# Fully integrated ultra-low power Kerr comb generation

Brian Stern[1,2], Xingchen Ji[1,2], Yoshitomo Okawachi[3], Alexander L. Gaeta[3], and Michal Lipson[2]

[1]School of Electrical and Computer Engineering, Cornell University, Ithaca, NY 14853, USA
[2]Department of Electrical Engineering, Columbia University, New York, NY 10027, USA
[3]Department of Applied Physics and Applied Mathematics, Columbia University, New York, NY 10027, USA
Corresponding Author: ml3745@columbia.edu

Optical frequency combs are broadband sources that offer mutually-coherent, equidistant spectral lines with unprecedented precision in frequency and timing for an array of applications[1-9]. Kerr frequency combs in microresonators require a single-frequency pump laser and have offered the promise of highly compact, scalable, and power efficient devices. Here, we realize this promise by demonstrating the first fully integrated Kerr frequency comb source through use of extremely low-loss silicon nitride waveguides that form both the microresonator and an integrated laser cavity. Our device generates low-noise soliton-modelocked combs spanning over 100 nm using only 98 mW of electrical pump power. Our design is based on a novel dual-cavity configuration that demonstrates the flexibility afforded by full integration. The realization of a fully integrated Kerr comb source with ultra-low power consumption brings the possibility of highly portable and robust frequency and timing references, sensors, and signal sources. It also enables new tools to investigate the dynamics of comb and soliton generation through close chip-based integration of microresonators and lasers.

Frequency combs based on chip-scale microresonators offer the potential for high-precision photonic devices for time and frequency applications in a highly compact and robust platform. By pumping the microresonator with a single-frequency pump laser, additional discrete, equidistant frequencies are generated through parametric four-wave mixing (FWM), resulting in a Kerr frequency comb[10,11]. Under suitable conditions temporal cavity solitons can be excited, which results in stable, low-noise combs with ultraprecise spacing[12–15]. Many applications require such tight frequency and timing stability, including spectroscopy[1–3], low-noise microwave generation[4,5], atomic clocks[6], lidar[7], and telecommunications[8,9]. Extensive research has explored different microresonator platforms to generate frequency combs for these applications[16–24].

While one of the most compelling advantages for microresonator combs is the potential for the pump source and the microresonator to be fully integrated, previous demonstrations using integrated resonators have relied on external pump lasers that are typically large, expensive, and power hungry, preventing applications where size, portability and low power consumption are critical. Power-efficient integrated lasers have been developed using silicon laser cavities with bonded or attached III-V gain sections[25–28], but losses in these silicon waveguides make comb generation impractical at low power. On the other hand, silicon nitride ($Si_3N_4$) microresonators were recently demonstrated with record low parametric oscillation thresholds[21] due to the high quality factors ($Q > 3\times10^7$), high nonlinearity ($n_2 \sim 2.4\times10^{-19}$ m$^2$/W), and small mode volume (ring radius $\sim$ 100 µm). Additionally, due to $Si_3N_4$'s high index of refraction ($n \sim 2.0$) and low loss, compact, tunable $Si_3N_4$ laser cavities with narrow linewidth have been demonstrated[29,30]. $Si_3N_4$ is a common complementary metal oxide semiconductor (CMOS)-compatible deposited material that can be fabricated at wafer scale, and the combination of efficient comb generation and available integration of active devices make it an ideal platform for complete integration of optical frequency combs.

Here we demonstrate a Kerr comb source on a fully integrated $Si_3N_4$ platform, using a compact, low-power, electrically-pumped source. In our approach (Fig. 1), a gain section based on a III-V reflective semiconductor optical amplifier (RSOA) is coupled to a $Si_3N_4$ laser cavity, which consists of two Vernier microring filters for wavelength tunability and a high-$Q$ nonlinear microresonator (Fig. 1b). The nonlinear microresonator serves two purposes. First, it generates a narrowband back-reflection due to Rayleigh scattering[16], effectively serving as an output mirror of the pump laser cavity, as we previously demonstrated[29]. Second, the microresonator generates a frequency comb through parametric FWM. In this way, the comb generation and pump laser are inherently aligned, a configuration that was previously explored using resonators in fiber laser cavities with fiber amplifiers[18,31]. Fully integrating the comb source allows the flexibility to use such a configuration, avoiding the typical chain of discrete components found in all previous Kerr comb demonstrations. Figure 1c shows the assembled millimeter-sized comb source, which has only electrical inputs and an optical output (see Methods for fabrication details).

We design the $Si_3N_4$ laser cavity to ensure tunable, single-mode lasing and provide sufficient pump output power for comb generation in the nonlinear microresonator. The lasing wavelength is controlled by the alignment of the two microring Vernier filters[27], which are in turn aligned with one of the modes of the larger microresonator shown in Fig. 1b. The filters' radii are 20 µm and 22 µm, corresponding to free spectral ranges (FSR) of 1.18 THz and 1.07 THz, respectively, which results in transmission at only a single frequency when the filters are aligned. Their resonance positions can be widely tuned using integrated resistive microheaters, as shown in Fig. 2a. The filters' transmission bandwidth is designed to have a full-width half-maximum (FWHM) of 15 GHz by ensuring strong coupling to the add and drop waveguides with a 5 µm coupling length. The optical gain in the laser cavity comes from electrical pumping of the III-V waveguide on the RSOA, which is coupled to the $Si_3N_4$ cavity on one end and strongly reflects at the opposite end (see Methods). The output coupler of the laser cavity is a 120 µm radius microresonator with a measured reflection of 40% on resonance (due to coupling between counter-propagating circulating beams resulting from Rayleigh scattering[29]), as shown in Fig. 2b. This level of reflection allows for high laser output power due to the high roundtrip gain of the RSOA. The measured transmission spectrum of the microresonator (Fig. 2d) corresponds to an intrinsic $Q$ of $(8.0 \pm 0.8) \times 10^6$. Based on this $Q$ and the anomalous group-velocity dispersion for the 730 x 1800 nm waveguide, simulations[32] indicate that a soliton-state frequency comb can be generated at 700 µW of pump power (Supplementary Fig. 1).

We show lasing with up to 9.5 mW output optical power using the integrated $Si_3N_4$ laser. In order to characterize the laser, we first operate the microresonator slightly detuned from resonance to ensure that only lasing occurs and a frequency comb is not generated. We observe lasing with over 60 dB side-mode suppression ratio (SMSR) (Fig. 2c). As shown in Fig. 2d, the lasing threshold is 49 mA, with a slope efficiency of 52 mW/A. The maximum on-chip output power of 9.5 mW is obtained at 277 mW (220 mA) electrical pump power. This corresponds to a 3.4% wall-plug efficiency. Additionally, we measure a narrow laser linewidth of 40 kHz using the delayed self-heterodyne method (see Methods). The relatively high output power and narrow linewidth is competitive with those of many bulk pump lasers, yet is significantly more compact.

Using our novel cavity design we generate a Kerr comb spanning 100 nm and achieve a mode-locked, single-soliton state with less than 100 mW electrical pump power consumption, enabling battery-operation applications. At 1.1 mW optical output laser power (corresponding to 78 mW electrical power), new frequency components begin to appear due to FWM in the microresonator. We then monitor the comb formation as the microresonator is tuned into the lasing mode's wavelength using a fixed electrical pump power of 130 mW. In order to generate the comb, the microresonator is roughly aligned with the filter rings such that lasing occurs at 1579 nm with 2.5 mW output power (Fig. 3a). As the microresonator is tuned into resonance, greater circulating power leads to comb formation, accompanied by high RF noise (Fig. 3b). Tuning the resonance further results in stable combs with smooth spectral envelopes characteristic of temporal cavity solitons. We measure a single-soliton state (Fig. 3c) with an associated drop in RF noise. Once generated, the soliton exhibits stable behavior without feedback electronics or temperature control, with no visible changes in the optical spectrum or output power until intentionally detuned. The signal-to-noise ratio of the central comb lines is approximately 50 dB, which is useful for spectroscopy applications[3]. We additionally show battery-operation of the comb source by supplying the pump current using a standard AAA battery. At 98 mW of electrical power from the battery, we measure 1.3 mW output optical power and a comb matching the single-soliton profile (Fig. 3d). These results represent unprecedented low power consumption for generating Kerr frequency combs and solitons with an integrated microresonator.

In order to show the versatility of this platform, we also demonstrate a more traditional but fully integrated configuration where the comb is generated in a microresonator that is different than the one used to generate the laser. In this second design, shown in Fig. 4a, the pump laser is distinct from the high-$Q$ microresonator. The Vernier filters and RSOA function the same as the first design, while a Sagnac loop mirror is included to serve as the output coupler with approximately 20% reflection. Since this mirror has a broadband reflection, tunable lasing can take place independent of the resonance position of the comb microresonator. With the microresonator fully off-resonance, we measure single-mode lasing at 1582 nm with 4.9 mW output power and over 60 dB SMSR (Fig. 4b) at 162 mW electrical pump power. By tuning the microresonator into resonance with the laser wavelength, we can generate a frequency comb spanning 110 nm (Fig. 4c). By further tuning into resonance such that the laser is effectively red-detuned[8], we observe a multiple-soliton state frequency comb spanning over 130 nm with the characteristic drop in RF noise (Fig. 4d). We model a two-soliton state comb and obtain a profile closely matching that of the experimental comb (Fig. 4d). The comb generation process in this second design is directly analogous to previous Kerr comb experiments[14,15], but this demonstration brings a high level of integration which affords greater flexibility in laser design and reduced power consumption.

This demonstration of the first fully integrated Kerr frequency comb source opens the door to countless new applications previously limited by the size and power requirements of discrete comb systems. The high level of integration enables new flexibility in designing the pump laser for generating the frequency comb, as evidenced by the two designs demonstrated here, consisting of an inherently aligned comb source enabled by a feedback reflection and a traditional modular configuration. Furthermore, the platform used here is CMOS-compatible for wafer-scale fabrication of robust, fully integrated photonic chips—potentially enabling wide deployment of precision references and sensors. The realization of a mode-locked Kerr comb on an integrated platform presents opportunities in many fields that rely on the precision and stability of frequency combs and solitons, including sensing, metrology, communications, and waveform generation. The low power consumption of the platform enables battery-powered and mobile systems, no longer relying on external lasers, movable optics, and laboratory set-ups.

## Methods

**Fabrication.** The $Si_3N_4$ devices are fabricated[21] by first growing 4 µm of $SiO_2$ on a crystalline silicon wafer using thermal oxidation to form the bottom cladding of the waveguides. Then 730 nm of $Si_3N_4$ is deposited using low pressure chemical vapor deposition (LPCVD). The wafer is annealed in two stages to remove hydrogen impurities. The waveguides are then patterned using electron beam lithography and etched using $CHF_3$ plasma etching. The waveguides are clad with 2 µm $SiO_2$. The microheaters are placed over the waveguides using 100 nm of sputtered platinum (with a titanium adhesion layer) and lift-off patterning.

**RSOA/$Si_3N_4$ Coupling and Electrical Connection.** The III-V RSOA gain chip used here is commercially available from Thorlabs (SAF 1126) and provides broad gain near 1550 nm. One side has 93% reflection and the other side is anti-reflection coated. This second side is coupled to the $Si_3N_4$ chip with the waveguides angled relative to the facets to further prevent reflections[29]. The $Si_3N_4$ chip is polished up to the end of a tapered 280-nm wide waveguide which is simulated to have less than 1 dB coupling loss to the mode of the RSOA waveguide. The two chips are attached and aligned using three-axis stages with micrometers. We measure an experimental 2 dB coupling loss. The RSOA is wirebonded to an electrical printed circuit board (PCB) for supplying the pump current from either a Keithley 2400 SourceMeter or a AAA battery with a tunable potentiometer. The Si$_3$N$_4$ chip's microheaters are connected to pads and interfaced with a DC wedge probe (GGB Industries) and controlled by a DAC (Measurement Computing). The Si$_3$N$_4$ waveguide output is formed as an inverse-taper to edge-couple to a lensed single-mode fiber.

**Laser Linewidth Measurement.** The laser linewidth is measured using the delayed self-heterodyne method[29]. The laser output is sent to an interferometer with one path delayed by 12 km of fiber (corresponding to a delay of 58 µs). The other path is phase modulated at 300 MHz. The resulting beat signal is measured on an electrical spectrum analyzer (Agilent E4407B) and a 40 kHz Lorentzian linewidth is determined.

**Acknowledgments**

We are grateful to Dr. Steven Miller, Chaitanya Joshi, Dr. Tong Lin, Dr. Utsav Dave, and Dr. Jae Jang for helpful discussions and to Mengjie Yu for help with soliton simulations. We also thank Min Chul Shin and Oscar Jimenez Gordillo for packaging advice. This work was supported by AFRL program award number FA8650-17-P-10 and the ARPA-E ENLITENED program (DE-AR00000843). This work was performed in part at Cornell NanoScale Facility, an NNCI member supported by NSF Grant ECCS-1542081.

**Figures**

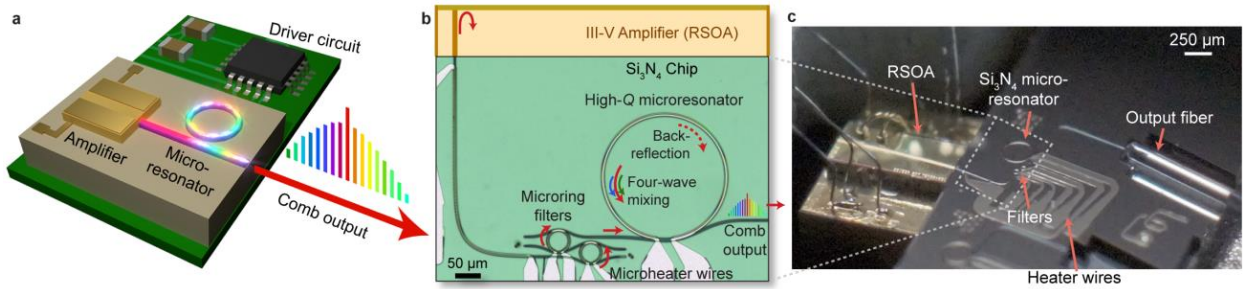

**Figure 1. Integrated frequency comb source.** (a) Concept illustration of an integrated Kerr comb source with an on-chip amplifier and microresonator. (b) Microscope image and diagram of integrated comb source, including laser cavity and nonlinear microresonator for comb generation. The amplifier waveguide provides electrically-pumped optical gain and includes a reflective facet on one end, while the opposite side is coupled to the $Si_3N_4$ portion of the laser cavity. The microring filters are tunable using integrated microheaters. The larger microresonator generates a partially reflected beam to form a second effective mirror of the laser cavity. This microresonator also has a high $Q$ to enable FWM and comb generation. (c) Photograph of the fully integrated comb source. The RSOA is edge-coupled to the $Si_3N_4$ chip and supplied with electric current via wirebonds, while the comb output is measured using an optical fiber.

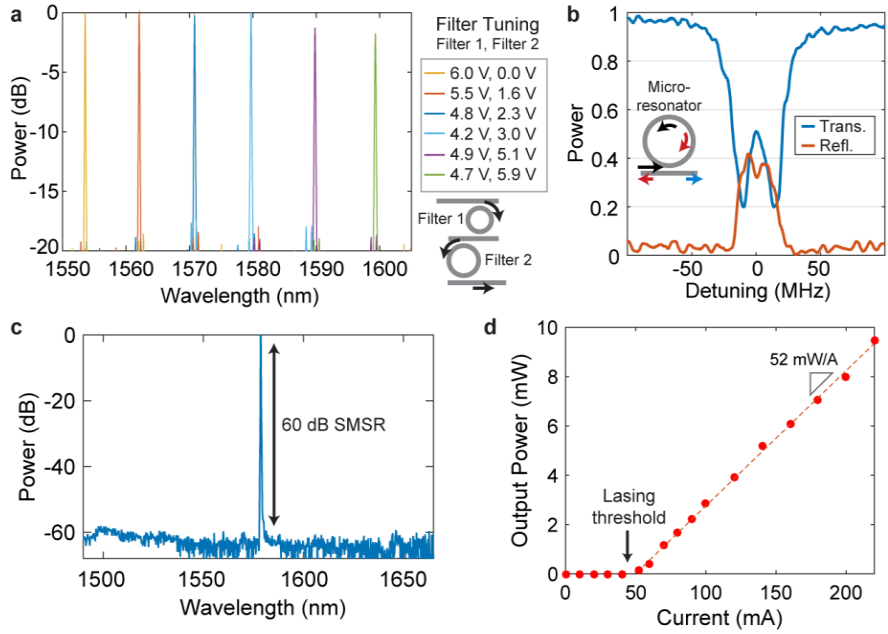

**Figure 2. Integrated III-V/Si₃N₄ laser characterization.** (a) Measured transmission spectrum for the Vernier filter rings. By adjusting the voltage applied to the microheaters, the filters relative detuning is adjusted and a single transmission wavelength is selected. (b) Measured optical transmission and reflection spectra (normalized) of the high-$Q$ microresonator. The 32-MHz resonance bandwidth reveals a $Q$ of $8 \times 10^6$. The narrowband reflection is generated by coupling via Rayleigh scattering between counter-propagating beams in the ring, which is apparent due to the resonance splitting observed from these degenerate beams. (c) Laser output spectrum showing single-mode lasing with over 60 dB side-mode suppression ratio (SMSR). (d) Output optical power of laser versus pump current at 1580 nm.

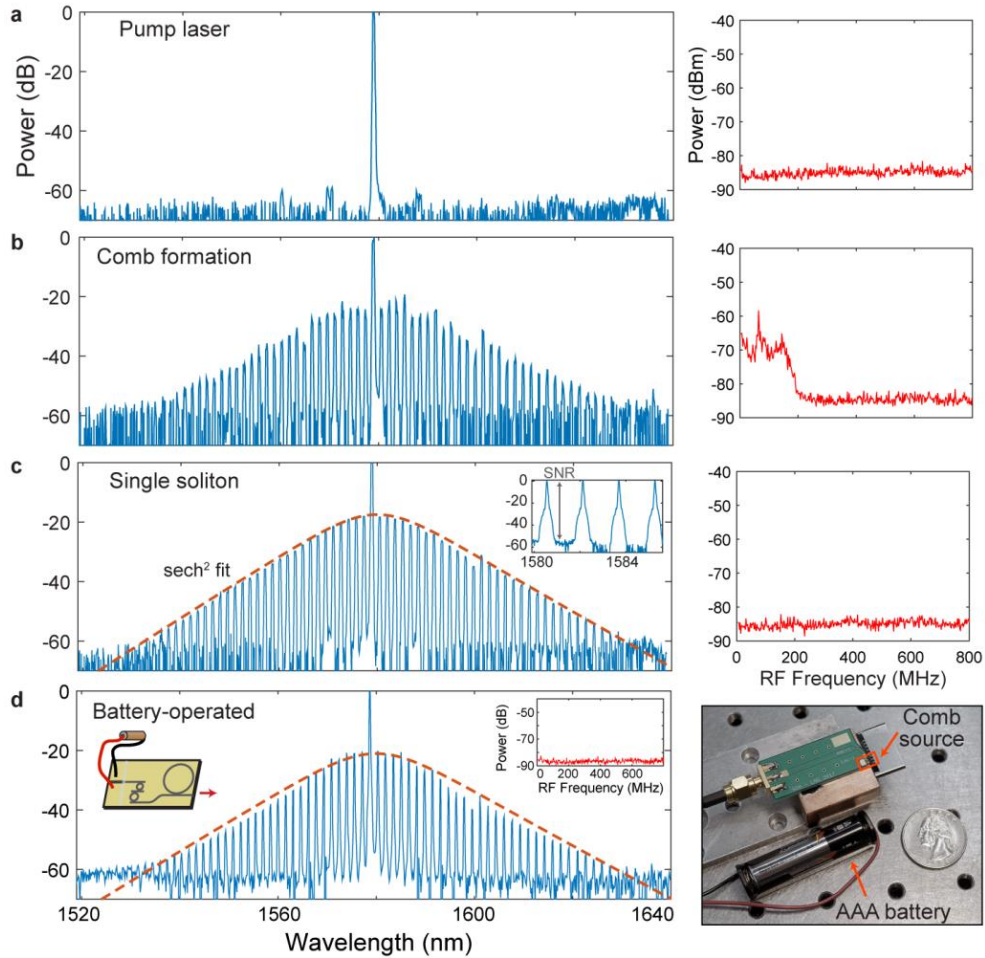

**Figure 3. Generation of mode-locked soliton frequency combs.** (a-c) Spectra of output from comb source as measured by an optical spectrum analyzer (OSA) at varying stages of comb generation with corresponding RF spectra. A current supply provides 130 mW electrical pump power. (a) Spectrum of laser output before tuning fully into resonance. Right: The RF noise is low since there is only single frequency lasing. (b) By tuning the microresonator resonance using the integrated microheater, a frequency comb forms. Right: Since the comb is not yet mode-locked, beating between different comb lines produces high RF noise below 200 MHz. (c) Single-soliton frequency comb achieved by tuning the microresonator such that the pump is slightly red-detuned from the resonance. The spectral envelope matches Fourier transform of the expected characteristic soliton sech profile (see Supplementary Fig. 2). Inset: The signal-to-noise ratio is approximately 50 dB (the laser linewidth is below the OSA resolution, but was separately measured to be 40 kHz). Right: The RF spectrum confirms the transition to a low-noise state. The resolution bandwidth used for all RF spectra is 100 kHz. (d) Frequency comb matching soliton profile generated with a AAA battery supplying pump power of 98 mW. Inset: RF spectrum showing low-noise state. Right: Photograph of integrated comb source with a printed circuit board and the battery.

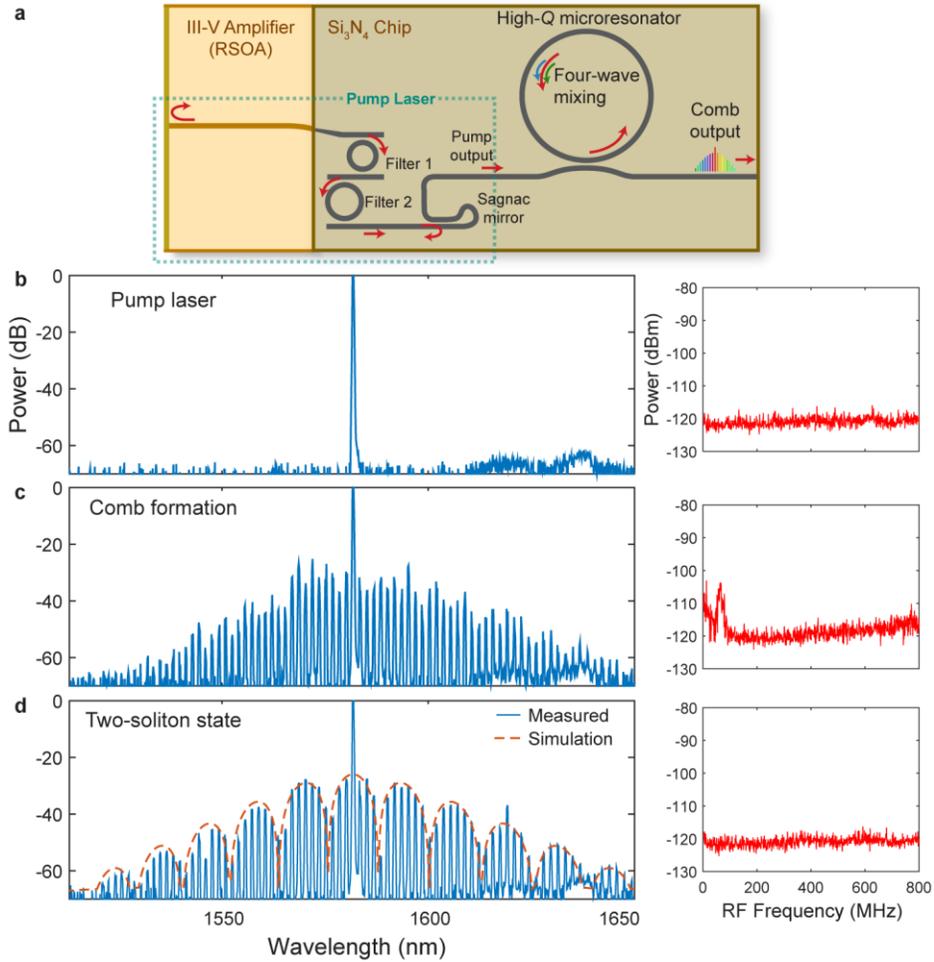

**Figure 4. Modular configuration for integrated comb source.** (a) Schematic of the modular comb source configuration, based on the typical external pump scheme. Here the integrated laser is distinct from the nonlinear microresonator, with a Sagnac loop mirror serving as the laser output coupler. (b-d) Optical output spectra at varying stages of comb generation with corresponding RF spectra. (b) Spectrum of laser output before tuning fully into resonance. The RF noise is low because there is only single frequency lasing. (c) By detuning the microresonator resonance using the integrated microheater, a frequency comb begins to form. The same is also achieved for red-shifting the pump wavelength relative to the microresonator resonance using the cavity phase shifter. Because this comb is not yet mode-locked, beating between different comb lines produces high RF noise. (d) Two-soliton frequency comb achieved by tuning the microresonator such that the pump is slightly red-detuned from the resonance. The RF spectrum confirms the low-noise state. The resolution bandwidth is 100 kHz.

**Supplementary Information**

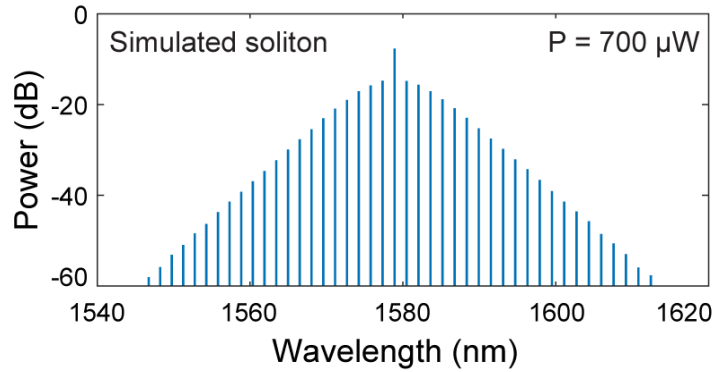

**Supplementary Figure 1. Comb generation simulation at low optical power.** Simulated optical spectrum of soliton generated with 700 µW optical pump power. The microresonator dimensions used in the model are 730 x 1800 nm with a radius of 120 µm, corresponding to a 194 GHz FSR (pulse repetition rate).

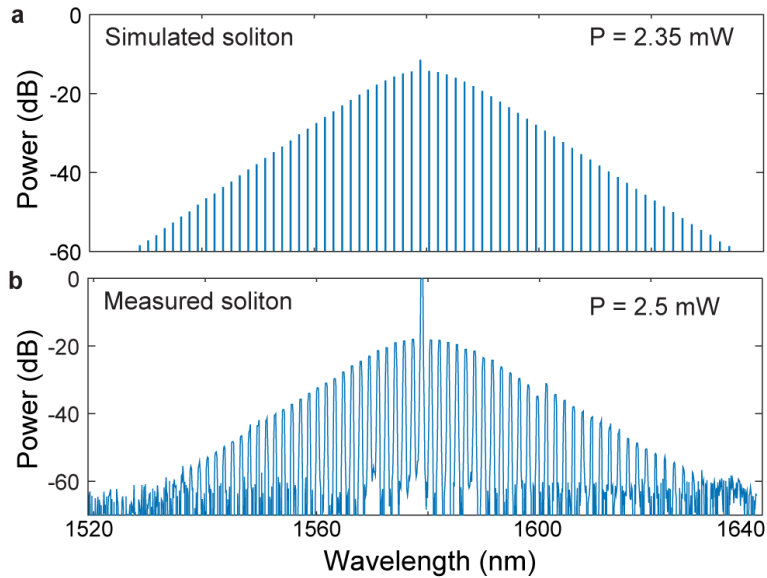

**Supplementary Figure 2. Comparison of simulated and measured solitons.** (a) Simulation of single-soliton comb generated with 2.35 mW pump power. The microresonator dimensions used in the model are 730 x 1800 nm with a radius of 120 µm, corresponding to a 194 GHz FSR (pulse repetition rate). (b) Optical spectrum of measured single-soliton comb (from Fig. 3c) generated with 2.5 mW pump power. The sech profile matches well with the simulated comb at slightly higher pump power.